# Fabrication and Characterization of a Starch-Based Nanocomposite Scaffold with Highly Porous and Gradient Structure for Bone Tissue Engineering

**Fereshtehsadat Mirab, Mohammadjavad Eslamian, and Reza Bagheri***

Polymeric Materials Research Group, Department of Materials Science and Engineering, Sharif University of Technology, Tehran, P.O. Box 11155-9466, Iran

*rezabagh@sharif.edu

**Abstract.** Starch based scaffolds are considered as promising biomaterials for bone tissue engineering. In this study, a highly porous starch/polyvinyl alcohol (PVA) based nanocomposite scaffold with a gradient pore structure was made by incorporating different bio-additives, including citric acid, cellulose nanofibers, and hydroxyapatite (HA) nanoparticles. The scaffold was prepared by employing unidirectional and cryogenic freeze-casting and subsequently freeze-drying methods. Fourier transform infrared (FTIR) spectroscopy confirmed the cross-linking of starch and PVA molecules through multiple esterification phenomenon in presence of citric acid as a cross-linking agent. Field emission scanning electron microscopy (FE-SEM) observations showed formation of aligned lamellar pores with a gradient pore width in the range of 80 to 292 µm, which well meets the pore size requirement for bone regeneration, and also well dispersion of cellulose and HA nanofillers within the scaffold matrix. Based on the mechanical testing results, the cellulose-HA reinforced scaffold possesses sufficient compressive modulus and yield strength for non-load bearing applications in the dry state; and also it presents fast responsive shape recovery in the wet state. According to in-vitro assessments, apatite phase mineralization was extensively induced in the presence of HA nanoparticles as heterogeneous nucleating sites. Also, it was revealed that cellulose and HA nanofillers decelerate and accelerate the scaffold biodegradation rate, respectively. MTT assay proved good cytocompatibility of the nanocomposite scaffold with osteoblast cells. Finally, it was shown that the introduced scaffold provides a suitable platform for the cells adhesion.

## 1- Introduction

Over the past few decades, bone tissue engineering using 3D porous scaffolds has been emerged as a promising approach for the treatment of bone defects [1]. In fact, the porous scaffold act as a template for bone tissue formation and also as a support for crucial cellular activities, such as cell attachment, proliferation, and differentiation [2]. An ideal scaffold for bone tissue regeneration must possess some requirements including appropriate porosity and pore size, surface roughness, sufficient mechanical property, biocompatibility, bioactivity, biodegradability [3]. Up to now, many researchers have tried to develop 3D bone scaffolds with tailored and biomimetic properties via introducing of new materials and processing techniques.

It is believed that the scaffold porosity higher than 90% and the presence of interconnected pores with minimum size of ∼100 μm are necessary for convenient cell migration and vascularization; moreover, the increase in the pore size (>300 μm) promotes the scaffold osteogenesis [4][5][6]. A highly porous scaffold containing aligned pore channels with gradient pore size can not only provide the abovementioned size requirements, but it also can enhance cell migration and nutrients transportation by capillary effect [7]. Such an appealing pore structure is achievable by unidirectional and cryogenic freezing of a hydrogel, aqueous suspension or solution, and subsequently, ice crystals removal via freeze-drying of the frozen sample [8][9][10]. Moreover, it is claimed that the surface roughness of the aligned channels, which is arisen from the dendritic growth of ice crystals during the cryogenic freezing, encourages cell adhesion, growth, and differentiation [11][12][13].

Recently, cryogel scaffolds based on natural polymers, especially polysaccharides, have attracted much attention because of their prominent biocompatibility and biodegradability [14][15]. Starch is a well-known, versatile, abundant, and inexpensive polysaccharide, which has exhibited a great potential as a promising biomaterial for bone tissue engineering [16][17][18]. However, since starch easily dissolves in the aqueous media, it shows low mechanical and shape stabilities in liquids. To compensate this drawback, the blending of starch with synthesized biopolymers has been previously proposed by some authors [19][20][21]. Among the all synthesized polymers, polyvinyl alcohol (PVA) has been extensively used in the biomedical field due to its excellent biocompatibility, water solubility, and degradability [15][22]. Moreover, the chemical cross-linking of starch and PVA molecules provides higher mechanical stability for the starch/PVA blend in fluids [22][23][24]. A high potential and favorable cross-linking agent is citric acid which is FDA approved for use in human body [25]. Citric acid is an organic acid with three carboxylic groups, and the cross-linking is made through esterification reaction between the carboxyl groups of citric acid and the hydroxyl groups of the PVA and starch molecules [26][22].

Polymeric scaffolds lack bioactivity and sufficient mechanical properties; hence, polymeric based nanocomposite scaffolds containing bioactive, biocompatible, and rigid fillers have been recently presented as new 3D scaffolds [27]. Cellulose is a crystalline, renewable and abundant natural polysaccharide, which is widely used for the reinforcement of the polymeric matrices [28]. Some authors have shown that incorporating cellulose nanofibers in starch-based scaffolds substantially improves their mechanical properties [29][30]. Moreover, it is reported that cellulose nanofibers diminish the biodegradation rate of starch-based scaffolds and exhibit no toxicity based on *in-vitro* evaluations [29]. Hydroxyapatite (HA) is another reinforcement which is widely used in polymeric-based scaffolds. In fact, HA is a rigid bioceramic whose composition is very close to the mineral component of natural bone; also it is well-known for its excellent bioactivity, biocompatibility and osteoconductivity [31][32]. Sundaram et al. [33] have shown that the addition of HA nanoparticles to a gelatin/starch based scaffold makes it much more stronger. Furthermore, it is believed that HA nanoparticles induce the apatite phase mineralization of polymeric based bone scaffolds, and can alter the degradation behavior of the polymeric matrix [34][3].

In this work, cellulose nanofibers and HA nanoparticles are incorporated to the cross-linked starch/PVA blend to form a novel biodegradable, biocompatible, bioactive, and water stable scaffold with a highly porous and gradient structure mimicking the bone extracellular matrix (ECM) architecture via unidirectional freeze casting followed by freeze-drying method. Then, the pore structure and mechanical and biological properties of the fabricated scaffold were investigated in detail. We believe that our proposed nanocomposite scaffold can be considered as a promising material for non-load bearing bone substitutes.

## 2- Experimental

### 2-1- Materials

Granular unmodified edible corn starch, 30 wt.% amylose and 70 wt.% amylopectin, was supplied by Glucosan Company (Iran). Polyvinyl alcohol ($M_w$= 72,000 gr/mole) and citric acid ($M_w$=192 gr/mole) were purchased from Merck. Mechanically synthesized cellulose nanofibers with the average diameter of 45 nm in gel state (containing 4 wt.% nanofibers) and synthesized hydroxyapatite nanoparticles (average diameter size ∼ 35 nm) were provided by Nano Novin Polymer Company and Pasture Institute (Iran), respectively.

**2-2- Fabrication of Scaffolds**

To study the structure and properties of the nanocomposite scaffold, three scaffolds containing different compositions were prepared as given in Table 1. The preparation procedure of the scaffolds is presented hereunder.

**Table 1.** The composition of the fabricated scaffolds

| Sample Code | Cumulative conc. of Starch and PVA | Starch/PVA ratio | Cellulose nanofibers concentration | | HA nanoparticles concentration | | Citric acid concentration | |
|---|---|---|---|---|---|---|---|---|
| | (% wt/v) | (wt/wt) | (pph) | (wt.%) | (pph) | (wt.%) | (pph) | (wt.%) |
| SP | 3 | 3:2 | 0 | 0 | 0 | 0 | 30 | 23.1 |
| SPC | 3 | 3:2 | 2.5 | 1.9 | 0 | 0 | 30 | 22.6 |
| SPCH | 3 | 3:2 | 2.5 | 1.5 | 30 | 18.5 | 30 | 18.5 |

* The pph values were considered as parts (by wt.) per hundred parts of the starch/PVA mixture in dry state.

**2-2-1- Preparation of starch/PVA solution**

In this study, 3% (wt/v) solution of starch/PVA (3:2 wt/wt) in deionized water was considered. To provide 100 ml of the solution, 1.2 gr PVA was dissolved in 50 ml deionized water at 90°C, and then was added to the mixture of 1.8 gr starch in 50 ml deionized water. Next, immersing in a 90 °C bath, the starch/PVA solution was mechanically stirred at 1500 rpm for 2 hrs, while its volume was kept constant at 100 ml by adding deionized water.

**2-2-2- Incorporating cellulose nanofibers**

2.5 pph cellulose nanofibers in gel state along with some deionized water were mixed for 30 min using a magnetic stirrer. Subsequently, for better dispersing of the nanofibers, the suspension was ultra-sonicated three times for 10 min. The ultra-sonicated suspension was added to the starch/PVA solution, which is obtained in section 2-2-1, and then mechanically stirred at 1500 rpm and 90 °C for 30 min. It should be noted, according to the literature, the nanofibers content should be lower than 3 wt.% to prevent fibers agglomeration. Therefore, the cellulose nanofibers content was set at 2.5 pph to get a well dispersion of the nanofibers within the matrix.

**2-2-3- Incorporating HA nanoparticles**

To break apart the as-received HA nanoparticle agglomerates, 30 pph HA nanoparticle was mixed with 5 pph citric acid and some deionized water, then the obtained suspension was ultra-sonicated three times for 10 min. Here, citric acid was employed to neutralize the surface charge of the HA nanoparticles, and thus to help with the dispersity of the nanoparticles in water [35]. Afterwards, the ultra-sonicated suspension was added to the ultimate suspension in section 2-2-2, and then mechanically stirred at 1500 rpm and 90 ℃ for 30 min. It should be mentioned that some investigators incorporated high amounts of HA particles (≥ 30 pph) to improve bioactivity of polymeric based scaffolds [33][36], but it has been reported that above 30 wt.% HA nanoparticles, the homogeneous dispersion of HA nanoparticles within the matrix would be difficult to achieve [37]. Hence, in this study, 30 pph of the HA nanoparticles was employed which is equall to 18.46 wt.% in the composite.

**2-2-4- Addition of crosslinking agent**

To prepare SP, SPC, and SPCH samples, the ultimate solution and suspensions in sections 2-2-1 to 2-2-3 were cooled to room temperature, and then 30, 30, and 25 pph citric acid was respectively added to them as crosslinking agent. It should be noted that since 5 pph citric acid had been previously consumed for the dispersion of HA nanoparticles (see section 2-2-3), here 25 pph citric acid was added to SPCH suspension. Finally, SP solution and SPC and SPCH suspensions were homogenized by mechanical stirring at 1500 rpm for 10 min at room temperature. It is noteworthy that it has been reported that the employment of high amount of citric acid (30 pph) not only extensively increases the cross-linking reaction but also makes no toxicity [26][22], therefore, we also used 30 pph citric acid in this work.

**2-2-5- Unidirectional freeze-casting followed by freeze-drying**

To unidirectionally solidify the prepared solution and suspensions, an apparatus described in the following was employed. The mold was a polytetrafluoroethylene (PTFE) tube ($\emptyset_{out} = 20$ mm, $\emptyset_{in} = 15$ mm, and $L = 6$ cm) which its bottom was sealed by an aluminum foil. The mold was insulated by a polystyrene foam, and placed on an aluminum substrate in contact with liquid Nitrogen. 5 ml of the solution/suspension was poured into the mold; and after freezing from bottom to top, the solidified sample was freeze-dried using an apparatus from Christ Company, Germany, under the following conditions: main drying at $-60$ °C and 0.005 mbar for 17 hrs, and final drying at $-76$ °C and 0.006 mbar for 31 hrs. Then, the achieved scaffold was ejected carefully from the PTFE tube, and located in an oven at 140 °C for 4 hrs to be thermally cross-linked. Finally, the fabricated scaffold was preserved in a silica gel desiccator.

### 2-3- Instrumental analyses

Fourier Transform Infrared (FTIR) spectra were obtained using ABB Bommem MB-100 spectrometer after preparing KBr-based pallets and recording five scan in the range of 4000 – 400 cm$^{-1}$. The surface morphology of the fabricated scaffolds was observed using TESCAN-Mira 3XMU Field Emission Scanning Electron Microscope (FE-SEM) equipped with an EDS detector. Prior to microscopy, the cross and longitudinal sections of the scaffolds were cryogenically cut and then coated with a thin gold layer. Compressive properties of the scaffolds in dry state were measured according to ASTM D638 using a Hounsfield H10KS universal frame equipped with 500 N load-cell at the cross-head speed of 1 mm/min and at ambient temperature.

### 2-4- Porosity measurement

The porosity of the nanocomposite scaffold was evaluated using gravimetric measurement [38] by means of Equation 1.

$$Porosity\ (\%) = \left[1 - \left(\frac{\rho_{app}}{\rho_0}\right)\right] \times 100 \qquad \text{(Equation 1)}$$

where $\rho_{app}$ is the apparent density of the porous nanocomposite scaffold, and $\rho_0$ is the density of non-porous nanocomposite. The reported porosity is the average of five repetitions.

### 2-5- In-vitro assays
### 2-5-1- Bioactivity and degradation tests

The bioactivity, i.e. the potential to form bonelike calcium phosphate layer, and degradation rate of the scaffolds were investigated within simulated body fluid (SBF) which contains ion concentrations similar to those found in human blood plasma. SBF solution was prepared based on Kokubo's instruction [39]. Subsequently, the fabricated scaffolds were soaked in SBF at 37 °C for different periods of time including 3, 7, 14, 21, and 28 days. After removal from SBF, the scaffolds were rinsed with deionized water and dried at 40 °C for 48 hrs under vacuum pressure. Possible formation of the apatite phase on the surface of the dried nanocomposite scaffold was monitored by FE-SEM; and also the degradation rate was calculated by means of Equation 2.

$$Dry\ Weight\ Loss\ (\%) = \frac{(W_0 - W_t)}{W_0} \times 100 \qquad \text{(Equation 2)}$$

where $W_0$ and $W_t$ are the dry weight of scaffold before and after degradation, respectively.

**2-5-2- Cytotoxicity assay**

The (3-[4,5-dimethylthiazol-2-yl]-2,5- iphenyltetrazolium bromide) MTT assay was conducted to assess the biocompatibility of the fabricated scaffolds. The extraction of scaffolds was carried out in accordance with the ISO 1993–5 protocol. The fabricated scaffolds were autoclaved, and then 0.1 g of them was incubated in 1 mL of culture medium at 37 °C for 3, 7, and 14 days. The culture medium was also preserved in similar condition and served as the negative control. The human osteoblast cell line (MG-63, National Cell Bank of Iran) was employed and cultured in Roswell Park Memorial Institute (RPMI) medium with 10% (v/v) fetal bovine serum (FBS-Serum, Germany), 100 U.ml$^{-1}$ penicillin, and 100 µg.ml$^{-1}$ streptomycin; and then it incubated at 37 °C in a humidified atmosphere containing 5% $CO_2$.

To evaluate the viability of the osteoblast cells, they were cultured into a 96-well microtiter plate (Nunc, Denmark) at a population density of 1×10$^4$ cells/well, and then incubated in the culture medium for 24 hrs. Then, the culture medium was discarded from the wells; and instead, 90 µl of the scaffold extract in the main concentration along with 10 µl FBS was added to each well. Again, after 24 hrs, the medium was removed and replaced with 100 µl of 0.5 mg/ml MTT (Sigma, USA) solution; and then the cells were incubated at 37 °C for 5 hrs. To solubilize the precipitated formazan crystals, 100 µl isopropanol (Sigma, USA) was added to each well, and subsequently, the plate was shaken for 20 minutes at 37°C. Finally, the optical density (OD) of the dissolved formazan was measured using a multiwall microplate reader (Stat Fax 2100, USA) at 545 nm, and the result was normalized with the negative control sample. In addition, to statistically analyze the obtained data, one-way ANOVA with LSD post hoc test using SPSS (version 16.0 SPSS Inc. Chicago IL) was utilized, and the statistical significance was considered at $p<0.05$.

**2-5-3- Cell adhesion assessment**

100 µl medium culture containing 30000-35000 MG-63 cells was poured on the surface of the sterilized nanocomposite scaffold. After incubating for 5 hrs, a specific amount of culture medium containing 10% FBS was added to the cell seeded scaffold, and the incubation was continued for 24 hrs. Afterwards, the solution on the scaffold was taken out; and then the scaffold was rinsed with phosphate-buffer saline (PBS) to remove non-adherent cells. The adhered cells were fixed with 3% (v/v) glutaraldehyde solution in PBS for 30 minutes at ambient temperature. Finally, the scaffold was gently washed with deionized water, dehydrated by graded alcohols (50%, 60%, 70%, 80%, 90%, 95%, and 100%, each for 5 min), and dried at room temperature. The dried scaffold was surface coated with gold nanoparticles, and viewed by FE-SEM.

**3- Results and discussion**

**3-1- Chemical structure evaluation using FTIR**

To establish whether the starch and PVA molecules are thermally cross-linked in the presence of citric acid, FTIR analysis was utilized for SP sample before and after heating at 140 ℃ for 4 hrs. The obtained FTIR spectra are presented in Figure 1a. Despite the difference in the absorption intensity of the peaks with the same wavenumbers, which is due to the difference in the amount of the analyzed samples, we can recognize two noticeable variations made by heating the sample.

First, a new absorption band is made around 1028 cm$^{-1}$ which is assigned to the formation of alkoxy-ester (C−O) bond attributed to the esterification of starch and PVA molecules with citric acid. As seen in Figure 1b, the esterification is occurred by reacting the carboxyl group of citric acid and the hydroxyl group of the starch or PVA molecule. Moreover, since a citric acid molecule has three carboxyl groups, citric acid could act as a crosslinking agent between the starch and PVA molecules through multiple esterification [22].

Second, the band around 3400 cm$^{-1}$, which is correlated to the stretching vibration of the OH groups, is broadened to the lower wavenumbers. This can be attributed to the OH groups form hydrogen bonds which link the molecules within the sample [40]. Consequently, we can conclude that the heat treatment induces the crosslinking of the starch and PVA molecules through the multiple esterification phenomenon and also the formation of more hydrogen bonds.

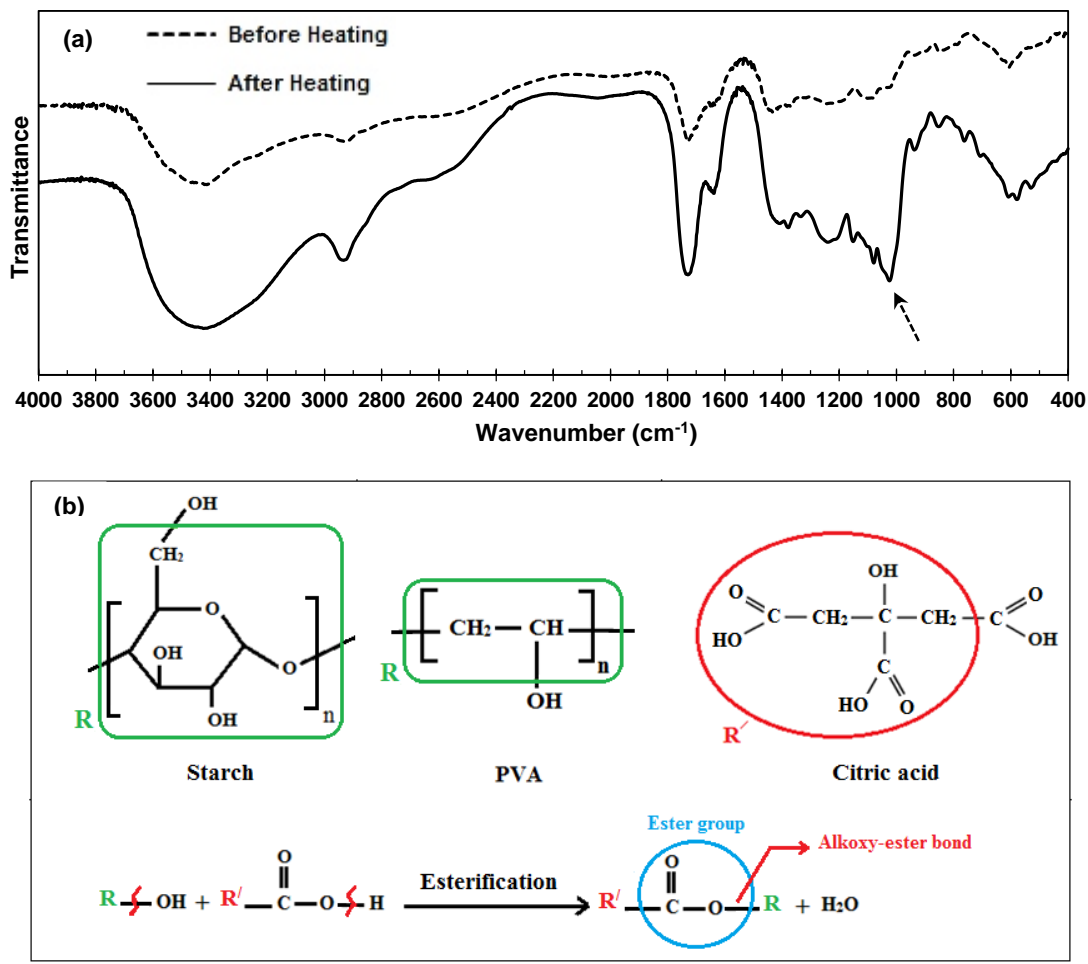

**Figure 1.** a) FTIR spectra of SP sample before and after heating at 140 °C for 4 hrs. The peak attributed to the alkoxy-ester bond is specified with an arrow. b) Esterification reaction between starch/PVA and citric acid.

**3-2- Investigation of pore structure and surface morphology**

Figure 2a shows the image of the fabricated SPCH scaffold. According to the gravimetric measurements, the porosity of the scaffold was 95%. This confirms the formation of a highly porous scaffold by the fabrication method used in this study. FE-SEM images of the lateral surface of the constructed scaffold, respectively from bottom to top, are illustrated in Figure 2b-2g. The observed pore structure is the replica of the ice crystals sublimated after freeze-drying. Indeed, the ice crystals morphology dictates the pore structure, so we can regulate the size and shape of pores by controlling processing parameters such as cooling rate and solution concentration. As seen, at the bottom of the scaffold (Figure 2b), fine cellular pores with an arbitrary array and maximum diameter of 50 μm are formed; but away from the bottom (Figure 2c-2g), aligned pore channels with gradient size are developed. A cross section image of the aligned pores is illustrated in Figure 2h. According to this figure, we can imply that the aligned pores are the replica of the lamellar ice crystals.

The average width of the aligned lamellar pores at the sections specified in Figure 2b-2g, are presented in Figure 2i. From section 1 to 5, the width of the aligned pores gradually increases within the range of 80 to 292 µm. This confirms the formation of aligned lamellar pores with gradient size in the fabricated scaffold. On the contrary, moving from section 5 to 6, the average width of the aligned pores decreases to 196 µm. This is due to high concentration of the remained solution at the final step of solidification. Also, by detailed scanning of Figure 2d, we can interpret the broadening mechanism of the aligned lamellar pores. As shown in this figure, some aligned pores are closed at the expense of the growth and widening of their adjacent pores. For instance, the six adjacent aligned pores demonstrated in this figure are converted to four wider adjacent pores by moving in the solidification direction. It is noteworthy that the pore sizes achieved in this study are in the range of those believed to be ideal for bone tissue engineering [41], and also the aligned channels can act as physical cues and guide the cells to migrate from one side of the scaffold to the other [13].

Based on all these observations, the formation mechanism of the ice crystals during unidirectional freezing of the solution can be schematically depicted as Figure 2j. In the region close to the cold substrate, owing to the tremendous nucleation and restricted growth rates of ice crystals, they form as equiaxed fine ice crystals. But moving away from the cold substrate, the presence of the unidirectional temperature gradient results in the establishment of the aligned lamellar ice crystals with peculiar anisotropic growth in the solidification direction [13]. As the aligned ice crystals grow, the distance between the solidification front and the cold source increases; therefore, supercooling at the solidification front diminishes and the aligned lamellar ice crystals tend to increase their thickness [42][43]. Moreover, due to the spatial constraint, the widening of aligned lamellar ice crystals is done at the expense of blocking of some aligned ice crystals during solidification.

The surface topography of the aligned pore channels is illustrated in Figure 2k. As seen, many branch-like features are arisen from the pore wall. The dendritic growth of the aligned ice crystals is responsible for creating of these features on the pore walls [13]. Another morphological feature is the bridge between the pore walls indicated by some arrows in Figure 2h. The bridge is originated from the entrapment of solutes within the ice crystals [44][43][43][45]. It is worth mentioning that the surface roughness made by the two above-mentioned features possibly could enhance attachment, proliferation and differentiation of anchorage- dependent osteogenic cells [5].

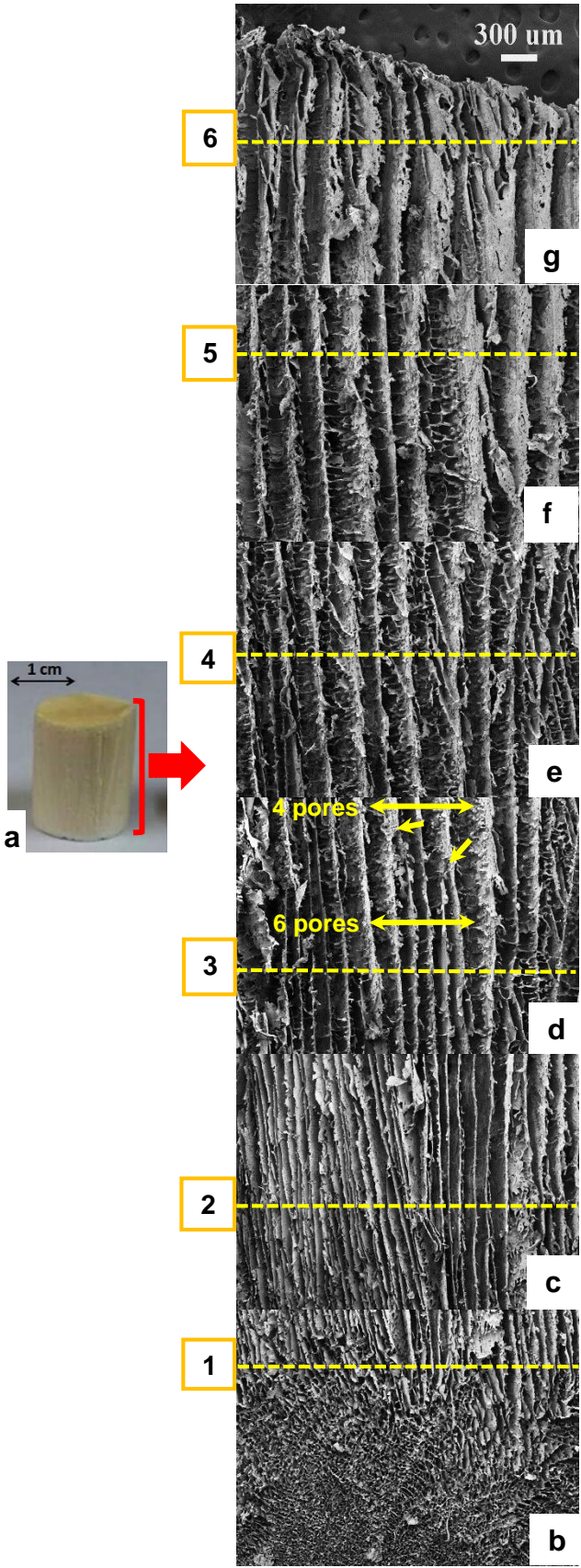
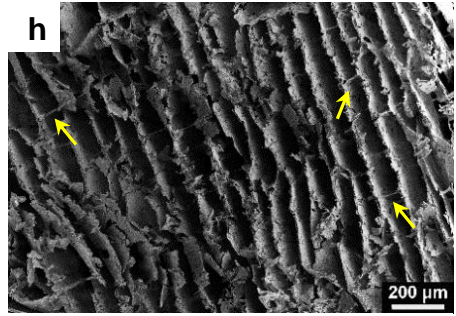
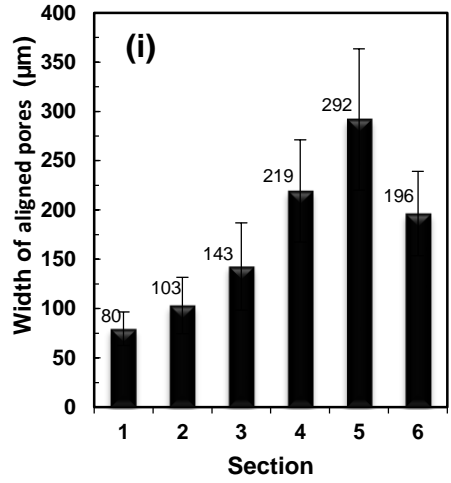
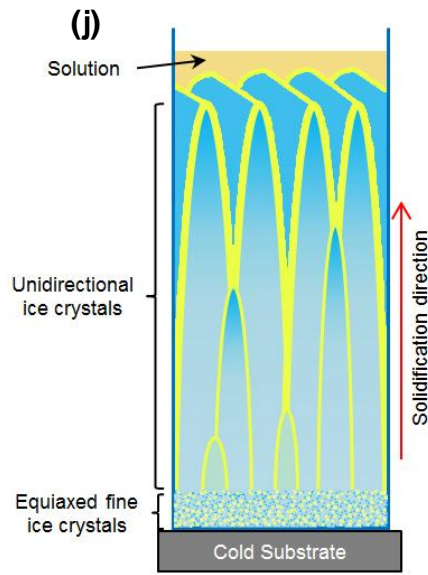
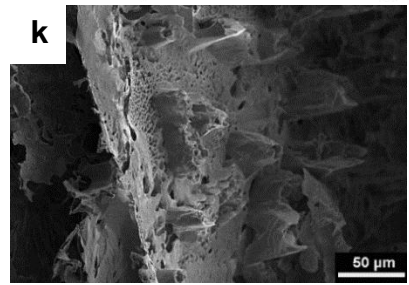

**Figure 2.** a) The fabricated SPCH scaffold. b, c, d, e, f, and g) FE-SEM images of the lateral surface of the scaffold respectively from bottom to top of the scaffold. h) FE-SEM image of the scaffold cross section perpendicular to the aligned pores direction. Arrows indicate some bridges between the aligned pore walls. i) The chart indicates the width of aligned pores at various sections specified in figures b, c, d, e, f and g. Twenty measurements was carried out for each section. j) The schematic indicates the formation and growth mechanism of ice crystals during solidification of the prepared solution. k) A FE-SEM image which shows the surface topography of the aligned pore walls.

### 3-3- Dispersion evaluation of cellulose and HA nanofillers

FE-SEM image of the surface of SPC scaffold is shown in Figure 3a. As seen, cellulose nanofibers have good dispersion within the matrix without any noticeable agglomeration. Observing the surface at higher magnification (Figure 3b), one can find that the dispersed nanofibers are well bonded to the matrix (see the areas specified by dashed circles). Similar nature of the cellulose nanofibers and the starch based matrix provides an excellent interfacial adhesion between them [30]. Diameter of the cellulose nanofibers available in Figure 3b is measured and presented as a histogram in Figure 3c. According to this histogram, the maximum diameter of the cellulose fibers is about 90 nm, which confirms the presence of nanometric cellulose fibers within the matrix. Moreover, the average diameter of the cellulose nanofibers is 48 nm which is close to the average diameter of the as-received nanofibers.

Figure 3d and 3e present the FE-SEM images of the surface of SPCH scaffold at two different regions. We can evidently observe that HA nanoparticles and cellulose nanofibers appropriately dispersed within the matrix. The EDS analysis of the nanoparticles is given in Figure 3f. According to the EDS graph, the peaks attributed to the phosphorus and calcium elements verify the presence of HA particles within the matrix. The diameter of the HA nanoparticles available in Figure 3d and 3e is presented as a histogram in Figure 3g. Based on this histogram, the average diameter of HA nanoparticles is 32 nm. Considering the average diameter of the as-received HA nanoparticles, we can find that the HA nanoparticles successfully dispersed within the matrix. It should be noted that the well dispersion of cellulose nanofibers and HA nanoparticles in the matrix indicates the suitability of the scaffold preparation method used in this study.

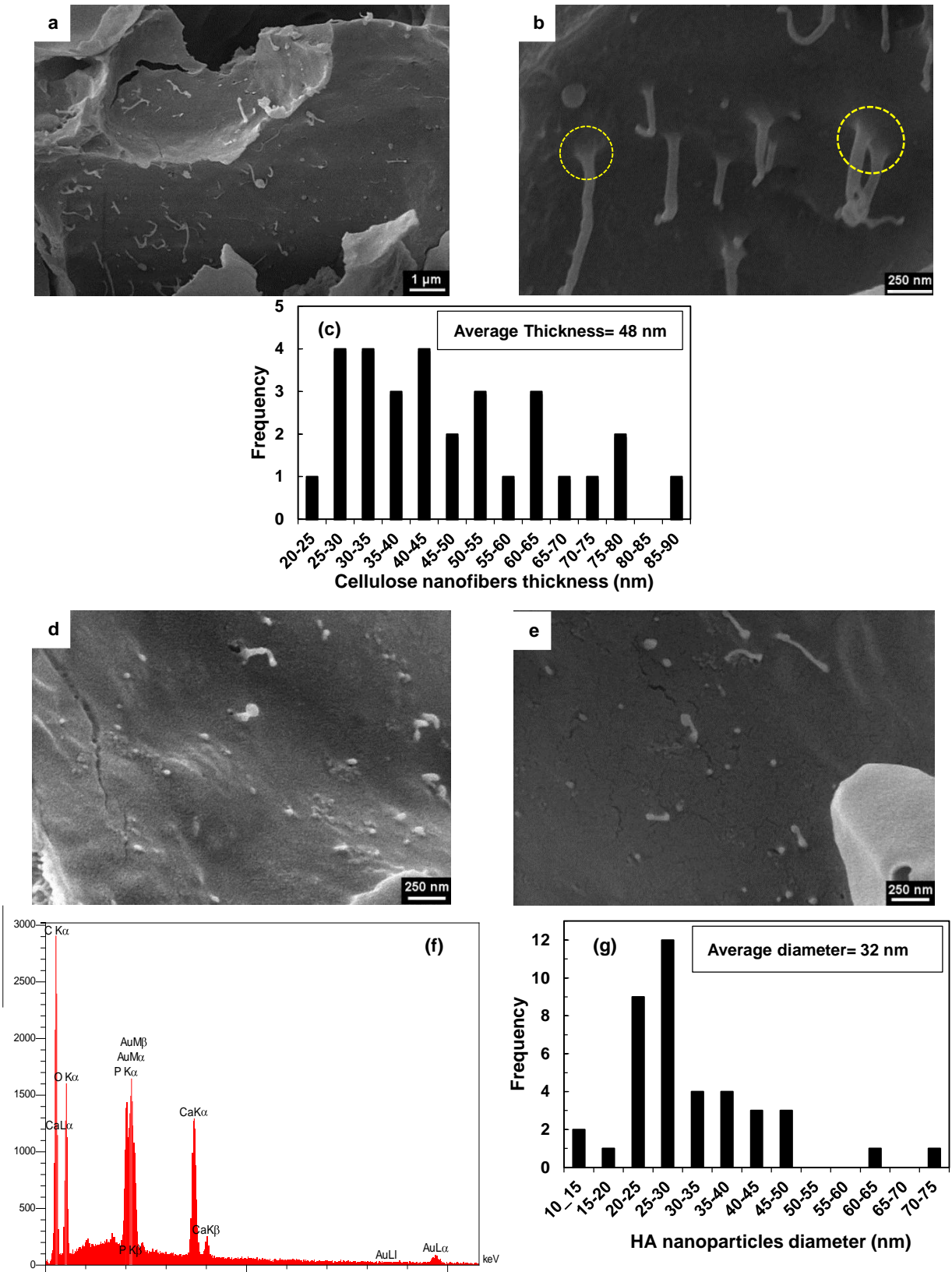

**Figure 3.** a, b) FE-SEM images of the surface of SPC scaffold at two magnifications: 18 KX (a) and 90 KX (b). The dashed circles exhibit the interfacial region between cellulose nanofiber and matrix. c) The histogram attributed to the diameter of the cellulose nanofibers available in image b. The histogram is based on 30 measurements. d, e) FE-SEM images of the surface of SPCH sample containing cellulose nanofibers and HA nanoparticles. f) The EDS graph of the HA nanoparticles available in Figure 3d. g) A histogram presents the diameter of HA nanoparticles seen in images d and e. This histogram is based on 40 measurements.

**3-5- Mechanical properties**

The compressive elastic modulus and yield strength (1% offset) of the fabricated scaffolds in dry state are presented in Figure 4a and 4b. As seen, SPC and SPCH scaffolds exhibit higher mechanical properties than SP scaffold. This implies that the cellulose nanofibers and HA nanoparticles have reinforced the scaffold matrix. In fact, the employment of rigid nanofillers which are well dispersed within the matrix and have a good adhesion to the matrix is responsible for the mechanical properties improvement. It should be noted that the obtained mechanical properties are in good agreements with those reported by other authors for highly porous polymeric based scaffolds applicable for non-load bearing bone defects [43][9][46]. For instance, Blaker et al. have fabricated a highly porous scaffold (~94% porosity) consisted of poly-D,L-lactic acid (PDLLA) filled with 15 vol.% Bio-glass particles; and they have reported the compressive modulus of 890±50 kPa and 1200±40 kPa respectively for the pure PDLLA and composite scaffold, and also a constant compressive yield strength (80±3 kPa) for both the pure and composite scaffolds [46].

Figure 4c shows the response of water-wetted SPCH scaffold under compression loading and unloading. As seen in this figure, the completely compressed scaffold recovers its initial shape after unloading. This phenomenon is referred to the shape-memory property of the scaffold. In fact, the crosslinking of starch and PVA molecules has led to the high elasticity of the scaffold. Furthermore, the diffusion of water into the intermolecular spaces triggers the shape recovery of the scaffold via reducing $T_g$ of the cross-linked matrix [47]. It is noteworthy that the shape recovery of the scaffold was carried out in seven seconds (For more intuitive observation, find the movie incorporated in the Supplementary Information section). Such a fast responsive shape recovery can be employed in minimal-invasive surgery with a compressed and small starting scaffold switching over to a larger structure in the body [48][23].

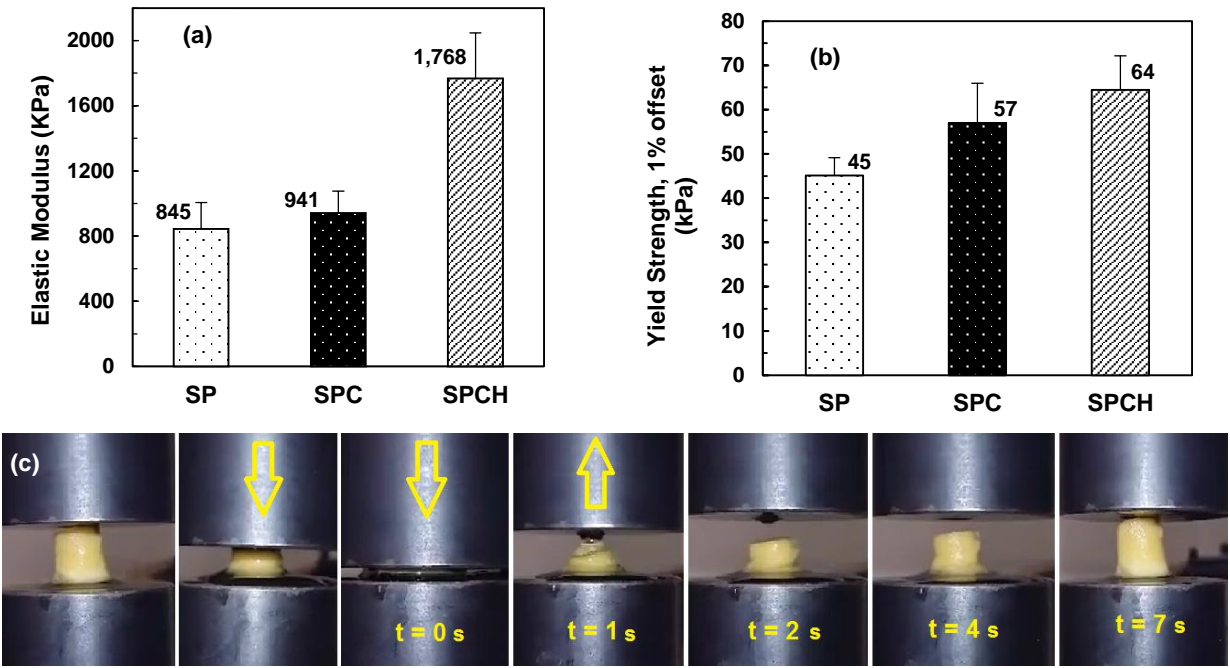

**Figure 4.** Compression properties of the fabricated scaffolds: a) Elastic modulus, b) Yield strength (1% offset). c) The water-wetted SPCH scaffold during compression testing: loading and unloading. The wet scaffold completely recovers its initial shape in seven seconds after unloading.

### 3-5- Biological evaluations

### 3-5-1- Mineralization

Figure 5 illustrates FE-SEM images of the surface of SPC and SPCH scaffolds after incubation in SBF for three days at 37 ℃. As seen in Figure 5a, the surface of the SPC scaffold is wrapped with a deposited layer which is attributed to the homogeneous formation of apatite layer on the scaffold surface in the presence of citric acid. Nourmohammadi et al. [49] have reported that the high chemical affinity of carboxyl groups in citric acid results in nucleation of apatitic calcium phosphate throughout the citrate starch based scaffold.

Now, looking at Figure 5b-5e, which show the surface images of the SPCH scaffold at different magnifications, one can clearly observe the mineralized apatite phase with the well-known cauliflower-like shape made of fine flake-like crystals. In this sample, HA nanoparticles act as nucleation sites and induce the heterogeneous nucleation of the apatite phase from the SBF which is a metastable calcium phosphate solution [34]. Comparing the surface images of the SPC and SPCH samples, it can be concluded that HA nanoparticles noticeably improve the apatite phase mineralization on the scaffold. In fact, the extensive formation of the cauliflower-like apatite phase on the entire surface of SPCH scaffold affirms its excellent bioactivity [50][3].

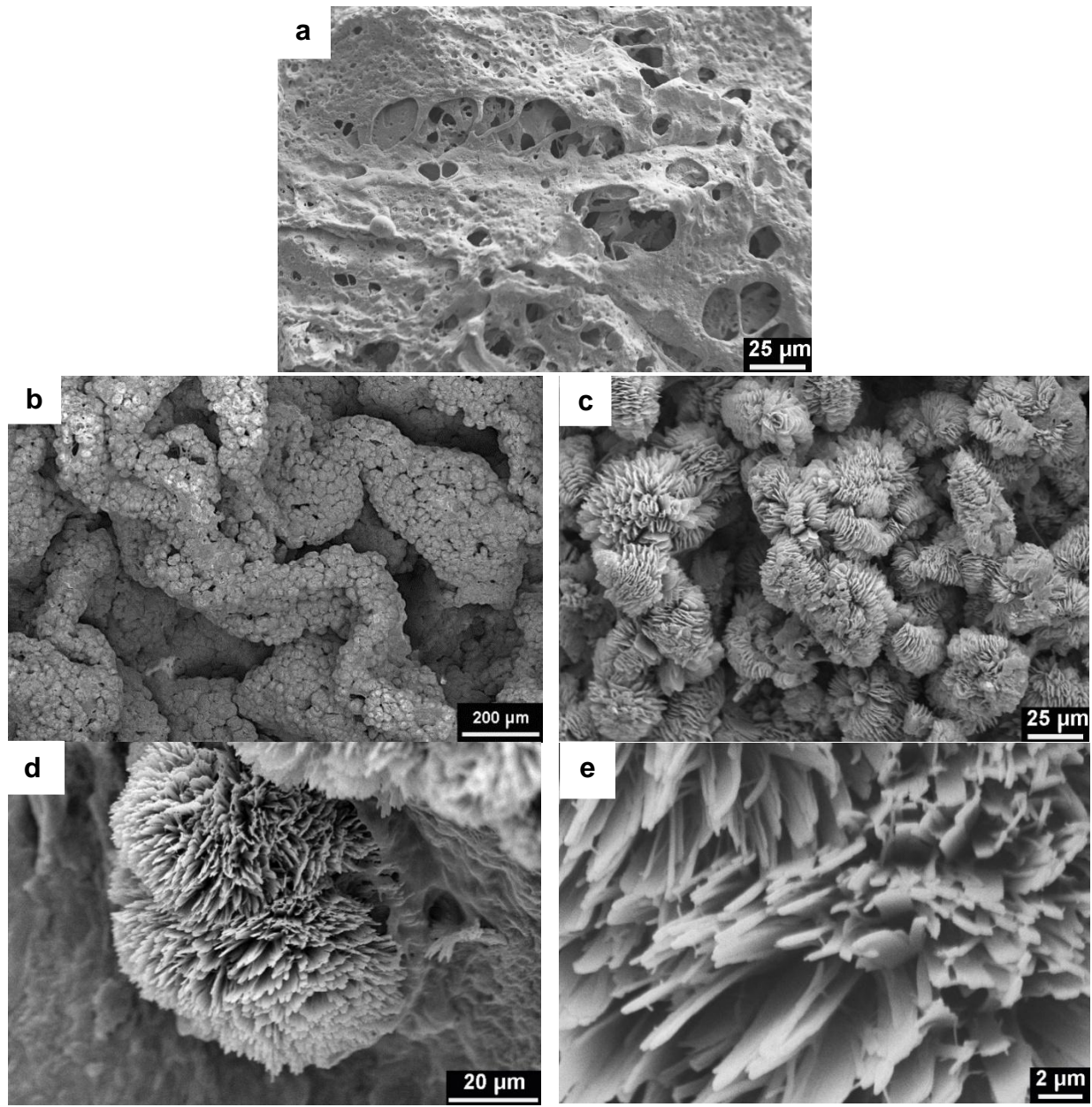

**Figure 5.** Apatite phase mineralization on SPC and SPCH scaffolds after incubating in SBF for three days at 37℃. a) FE-SEM image from the surface of SPC scaffold represents the homogeneous formation of apatite layer on the scaffold. b-e) FE-SEM images from the surface of SPCH scaffold show the heterogeneous formation of cauliflower-like apatite phase made of flake-like fine crystals on the scaffold at different magnifications: b) 150 X, c) 500 X, d) 1 KX, and e) 5 KX.

**3-5-2- Biodegradation**

Figure 6 indicates *in vitro* biodegradation behavior of the fabricated scaffolds based on dry weight loss percentage after incubating in SBF for 3, 7, 14, 21, and 28 days. As seen, SP and SPC scaffolds are increasingly degraded through time, and the degradation rate of SPC is noticeably lower than SP. The final weight loss of SP and SPC scaffolds over 28 days of exposure in SBF respectively are 27.1% and 4%. The very low degradation rate of SPC scaffold is ascribed to the formation of strong hydrogen bonds between hydroxyl groups of the rigid cellulose nanofibers and starch/PVA molecules [29]. The degradation behavior of SPCH scaffold shows that the addition of HA nanoparticles to the scaffold considerably builds up its degradability in each interval. As seen, the weight loss percentage of SPCH scaffold after 3 days incubation in SBF is 31.9%; and then it increases up to 45% after incubating in SBF for 14 days. As it is claimed by other authors, the interfacial region between the HA nanoparticles and the matrix are very susceptible to hydrolysis; hence, the employment of HA nanoparticles raises the degradation rate of the scaffold by accelerating its hydrolysis process [51][3]. In addition to the increase in the scaffold degradation kinetic, the HA nanoparticles simultaneously cause mineralization of the apatite phase on the scaffold in SBF medium. The decrement in the dry weight loss of SPCH scaffold after 21 and 28 days incubation in SBF can be attributed to the extensive formation of the apatite phase on the scaffold.

In a word, cellulose nanofibers and HA nanoparticles alter the degradation behavior of scaffold by decelerating and accelerating the degradation rate, respectively. Likely, by varying the amounts of cellulose nanofibers and HA nanoparticles, we might be able to control the degradation rate of the nanocomposite scaffold based on our desired function. A firm conclusion, however, needs further studies.

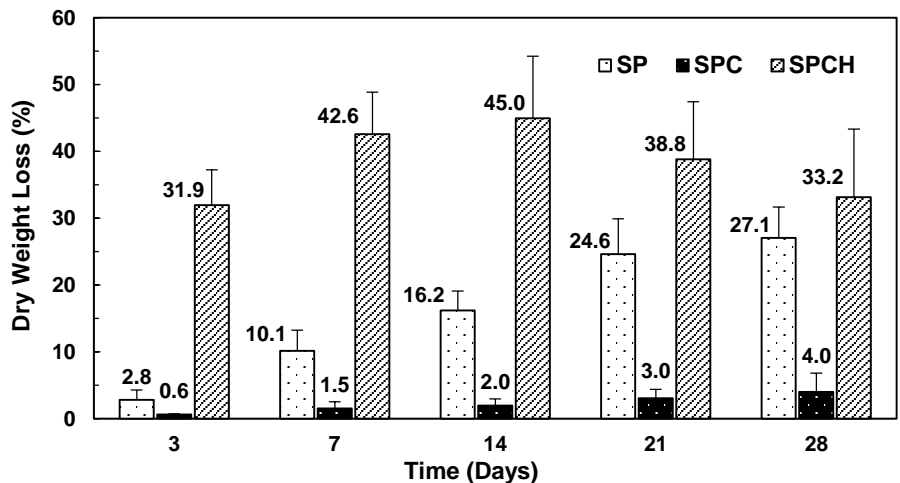

**Figure 6.** Biodegradation behavior of the fabricated scaffolds after incubating in SBF at 37 ℃ for different periods of time. The data are given based on dry weight loss of the scaffolds and are the average of three measurements.

### 3-5-3- Cytocompatibility

According to MTT assay, the viability percentage of MG-63 osteoblast cells in contact with the scaffolds' extraction obtained after different extraction times are presented and compared with the control samples in Figure 7. As seen, the cellular viability of all samples is more than 94%. This shows that the fabricated scaffolds have good cytocompatibility with the human bone cells, which is ascribed to the well biocompatibility of the scaffolds constitutes and the green process of the scaffolds fabrication. The little decrement in the cell viability of SPC scaffold can be attributed to the presence of some acidic compounds within the purchased gel containing cellulose nanofibers used for fabrication of the scaffold. In fact, the acidic compounds negatively affect the viability of the cells by reducing the pH of culture medium. However, the employment of HA nanoparticles in SPCH scaffold compensates the pH drop by releasing some negatively charged ions to the medium, and thus builds up the cellular viability. It is worthwhile mentioning that the utilization of citric acid as crosslinking agent has not made any toxicity for the fabricated scaffolds.

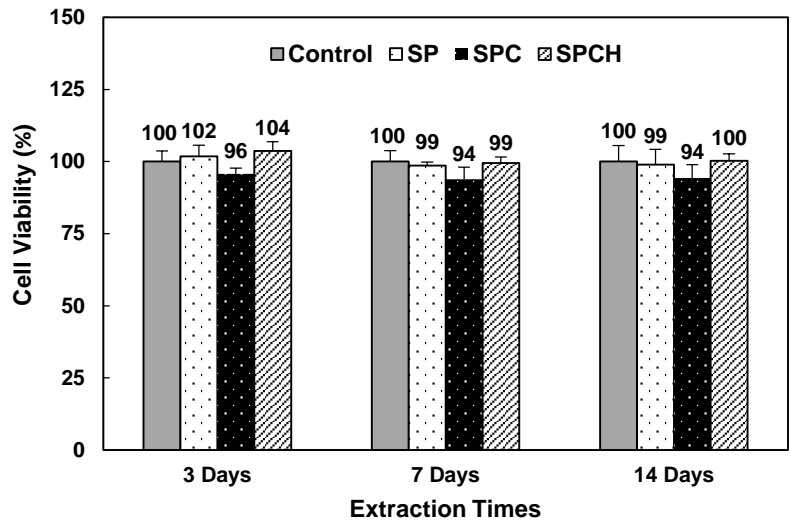

**Figure 7.** The viability (%) of MG-63 osteoblast cells in contact with the scaffolds' extraction obtained by MTT assay after different extraction times including 3, 7, and 14 days. Pure culture medium was employed as control sample.

### 3-5-4- cell adhesion

Figure 8 illustrates the FE-SEM images at different magnifications of the MG-63 osteoblast cells with round-shape morphology seeded on SPCH scaffold after 24 hrs. The images are taken from the interior of the scaffold, which show the migration of cells inside the pore channels and their spread on the scaffold matrix (Figure 8a). Good adhesion of the cells to the surface of the scaffold is evidenced in Figure 8b and 8c. The favorable interaction between the osteoblast cells and SPCH scaffold confirms the biocompatibility of the used materials in this study, and proposes that the fabricated porous structure could provide a suitable platform for migration, growth and proliferation of the cells.

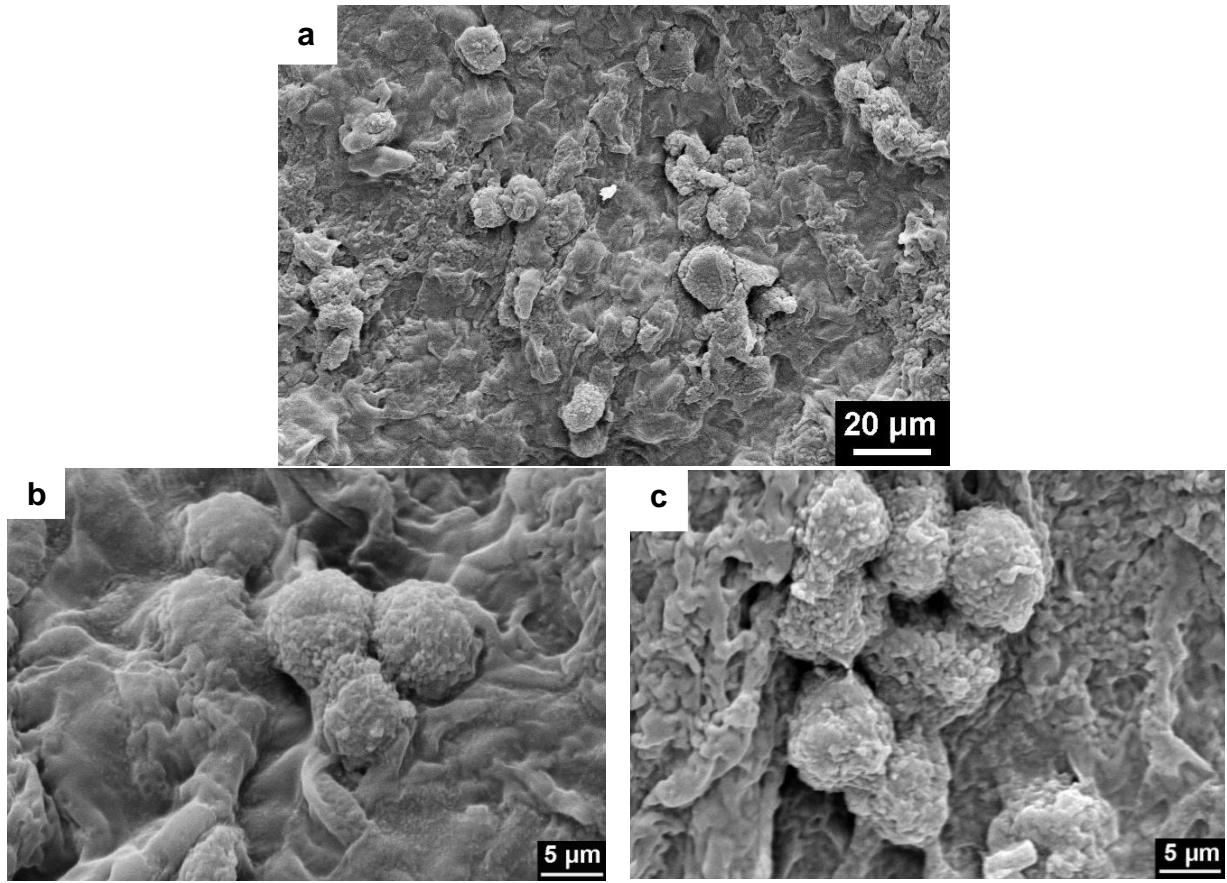

**Figure 8.** FE-SEM images from the interior of SPCH scaffold after osteoblast cell seeding for 24 hrs at different magnifications: a) 500X, b and c) 2500X. The round-shape osteoblast cells are well adhered to the matrix.

**Conclusion**

In the present study, a highly porous starch-based nanocomposite scaffold containing cellulose nanofibers and hydroxyapatite nanoparticles was fabricated via unidirectional freezing followed by freeze drying. To increase stability of the scaffold in aqueous media, starch was blended with PVA followed by cross-linking using citric acid. FTIR analysis confirmed that starch and PVA are thermally cross-linked through multiple esterification phenomenon. Porosity of the fabricated scaffold was 95% which verifies the formation of a highly porous scaffold. The pore structure of the scaffold from bottom to top contains cellular pores with maximum diameter of 50 µm and aligned pore channels with gradient width in the range of 80 to 292 μm, which could be appropriate for bone tissue regeneration. It was revealed that both cellulose nanofibers and HA nanoparticles, which are well dispersed and adhered to the matrix, improve the compressive properties of the scaffold. In addition, the water-wetted scaffold presents shape memory behavior after compression unloading. Extensive mineralization of apatite phase on the scaffold after incubating in SBF proved the high bioactivity of the scaffold. *In vitro* biodegradation assessment demonstrated that while cellulose nanofibers decelerate degradation rate of the scaffold, HA nanoparticles accelerate it. MTT assay ascertained excellent cytocompatibility of the scaffold with MG-63 human osteoblast cells. In addition, adhesion of the osteoblast cells to the scaffold surface confirmed their favorable interaction. In brief, the fabricated nanocomposite scaffold with gradient pore structure can be considered as a new biocompatible, biodegradable, bioactive, and mechanically fast responsive scaffold for minimally invasive filling of non-load bearing bone defects.